\documentclass[twocolumn]{aastex63}
\usepackage{amsmath}

\usepackage{graphicx}
\usepackage{mathtools}
\usepackage{amsmath,amssymb}
\usepackage{float}
\usepackage{xspace}
\usepackage{booktabs}
\usepackage[T1]{fontenc}
\usepackage[utf8]{inputenc}

\received{Received}
\revised{Revised}
\accepted{Accepted}
\submitjournal{ApJ}

\graphicspath{{./}{figures/}}

\begin{document}

\title{Accelerating Radiative Transfer for Planetary Atmospheres by Orders of Magnitude with a Transformer-Based Machine Learning Model}

\correspondingauthor{Isaac Malsky}
\email{isaac.n.malsky@jpl.nasa.gov}

\author[0000-0003-0217-3880]{Isaac Malsky}
\affiliation{Jet Propulsion Laboratory, California Institute of Technology, Pasadena, CA 91109, USA}

\author[0000-0003-3759-9080]{Tiffany Kataria}
\affiliation{Jet Propulsion Laboratory, California Institute of Technology, Pasadena, CA 91109, USA}

\author[0000-0003-1240-6844]{Natasha E. Batalha}
\affiliation{NASA Ames Research Center, Moffett Field, CA 94035, USA}

\author[0000-0002-3168-0139]{Matthew Graham}
\affiliation{California Institute of Technology, Pasadena, CA, 91125, USA}

\begin{abstract}
Radiative transfer calculations are essential for modeling planetary atmospheres. However, standard methods are computationally demanding and impose accuracy-speed trade-offs. High computational costs force numerical simplifications in large models (e.g., General Circulation Models) that degrade the accuracy of the simulation. Radiative transfer calculations are an ideal candidate for machine learning emulation: fundamentally, it is a well-defined physical mapping from a static atmospheric profile to the resulting fluxes, and high-fidelity training data can be created from first principles calculations. We developed a radiative transfer emulator using an encoder-only transformer neural network architecture, trained on 1D profiles representative of solar-composition hot Jupiter atmospheres. Our emulator reproduced bolometric two-stream layer fluxes with mean test set errors of $\sim$1\% compared to the traditional method and achieved speedups of more than 100x. Emulating radiative transfer with machine learning opens up the possibility for faster and more accurate routines within planetary atmospheric models such as GCMs. 
\end{abstract}

\section{Introduction}\label{sec:Introduction}
At the heart of almost every computational model of an exoplanet atmosphere lies a radiative transfer routine that determines how radiation is scattered, absorbed, and emitted as it propagates through the atmosphere. These methods are computationally demanding, as they require solutions to integro-differential equations in many distinct wavelength bins. Computational cost forces simplifications of the physics implemented within atmospheric models, lowers model temporal and spatial resolutions, and restricts the number of models that can be run in grid searches.

Three-dimensional General Circulation Models (GCMs) are a class of models for which the computational cost of radiative transfer is a major barrier. At their core, GCMs self-consistently simulate exoplanet atmospheres using a chosen set of fluid dynamics and radiative transfer equations. GCMs are critical to the field, as they allow for the characterization of inherently 3D exoplanet atmospheres. However, completing a single GCM run can take months of real-world time \citep[e.g.,][]{showman2002, showman2009, Amundsen2016, Drummond2018}.

To illustrate why radiative transfer is so computationally demanding, consider an example use case within a GCM. The radiative transfer equations must be solved in order to get the instantaneous heating rates across the entire global structure, and recalculated throughout the model evolution. Most GCMs divide the model into thousands of 1D columns and solve the multiwavelength radiative transfer equations independently, under the plane-parallel assumption \citep[e.g.,][]{Heng2017}. Atmospheric optical properties vary many orders of magnitude with wavelength, and must be calculated for many different ``bins'' \citep[e.g.,][]{showman2009, kataria2015, lee2021}. A typical GCM has on the order of 100 spectral bins (wavelength bins and Gauss points), 50 vertical layers, and 2500 latitude and longitude grid points. From these 1D columns, layer fluxes and heating rates are determined based on the static state at each timestep. Throughout a single model simulation, fluxes must be computed for every spectral bin, at each vertical level, across both horizontal dimensions, and at approximately every model timestep. Depending on the spatial, temporal, and wavelength resolution of the model, this can result in many billions of operations.

Every GCM introduces simplifications for each physical process due to computational limitations. For example, gas opacities can be represented through two constant absorption coefficients \citep{Roman2017,Roman2019}, the k-distribution method with the correlated-k approximation \citep{Goody1989, Lacis1991, Fu1992, Amundsen2016}, or the full line-by-line sources \citep{ding2019}. Similarly, the radiative transfer equation can be solved with varying approximations, such as the absorption approximation or variational iteration method for thermal radiation \citep{he1999, zhang2017vim, lee2024}. One common method in exoplanet GCMs is the two-stream approximation with multiple scattering \citep{Toon1989, Heng2018} to calculate fluxes, coupled with a k-distribution scheme (with a minimal number of bins for computational feasibility) to calculate optical properties. More complex solutions, however, can more accurately account for the angular dependence of radiative transfer \citep[e.g.,][]{Stamnes1988, Rooney2023, Rooney2024}. \cite{lee2024} compared several methods of calculating infrared fluxes with scattering to benchmark 64-stream DISORT solutions (Discrete-Ordinate-Method Radiative Transfer; \citealt{Stamnes1988}), finding typical errors around $\sim$1\%. However, they find substantially larger errors ($\gtrsim$50\%) at pressures below 1 bar when clouds are also present. Multi-stream solutions are becoming essential for accurately modeling scattering environments (e.g., cloudy atmospheres) and interpreting high-fidelity observations from JWST and next-generation telescopes \citep[e.g.,][]{Lee2025}.

% Recently, Emulators
Recently, a variety of machine learning methods have emerged to emulate physical processes, include more complex radiative transfer into GCM calculations \citep[e.g.,][]{Huppenkothen2023}, and have been incorporated into retrieval frameworks \citep{Marquez2018, Zingales2018, Cobb2019, Nixon2020, Santiago2022, Vasist2023, Duque2025}. Machine learning models iteratively adjust internal learned weights based on a set of examples (the training data) in order to ``learn'' how to approximate the input-output solutions. By approximating the underlying physical relations present within training data, these models effectively skip computationally intensive intermediate steps, directly predicting the desired outputs from physical inputs. Machine learning emulation of radiative transfer has already shown promise in Earth and Solar System GCMs. For example, \cite{Ukkonen2021} showed that a recurrent neural network (RNN) was able to emulate fluxes within an Earth atmosphere model to within 1\%, and \cite{Tahseen2024} showed that a similar RNN produced more than a 100-fold speed up over traditional radiative transfer methods in a Venusian GCM.

In this work, we developed a machine learning model capable of emulating radiative transfer in 1D atmospheric columns, tailored for hot-Jupiter GCMs (but easily expandable for a wider diversity of planets, or different atmospheric model types). The goal of the model was to take as input 1D atmospheric profiles (with varying pressure-temperature profiles, incident fluxes, and other parameters), and to output bolometric net layer fluxes. A successful model is one that matches the training data input-output radiative transfer solutions with minimal error, but at a fraction of the computational cost of the traditional method. In Section \ref{sec:Methods}, we outline the details of the 1D profile creation, forward model generation, and our transformer-based emulator training. In Section \ref{sec:results}, we evaluate the accuracy of the model and quantify the computational cost. We summarize the results, identify limitations of the model, and discuss future avenues for expanding the model in Section \ref{sec:discussion}.

\section{Methods}\label{sec:Methods}
We use the word emulator (also called a surrogate model) to describe a machine-learning model that approximates the input-to-output mapping of a classical, deterministic, physical solver. For example, traditional methods for radiative transfer produce fluxes from a set of input conditions (i.e., optical properties, temperature). By training on data generated using traditional methods, a machine learning emulator can predict fluxes, given the set of input conditions, thereby emulating the results of the original solver. Although these models will only ever approach the accuracy of the training data, they can provide orders-of-magnitude speedups over the traditional methods.

This work is novel in two ways. First, we adopted a modular framework for creating the radiative transfer training data and machine learning model. \cite{Tahseen2024} developed a machine learning radiative transfer emulator for the OASIS GCM \citep{Mendonca2015}. They showed high accuracy and a 147x speed up over traditional radiative transfer methods. However, this emulator was trained on the 1D profiles and radiative transfer results from the OASIS model. Here, rather than directly training the model using outputs from a specific GCM, we leveraged a standalone 1D forward model. This modular approach allows for a greater degree of flexibility in creating the training set and choosing the methods with which we calculate the radiative transfer. Second, we used an encoder-only transformer architecture \citep{Vaswani2017}. This type of model is particularly well-suited for sequence regression tasks with complex non-linear relationships. To the authors' knowledge, this is a new approach to calculating atmospheric radiative transfer.

\begin{figure*}
\begin{center}
\includegraphics[width=0.8\textwidth]{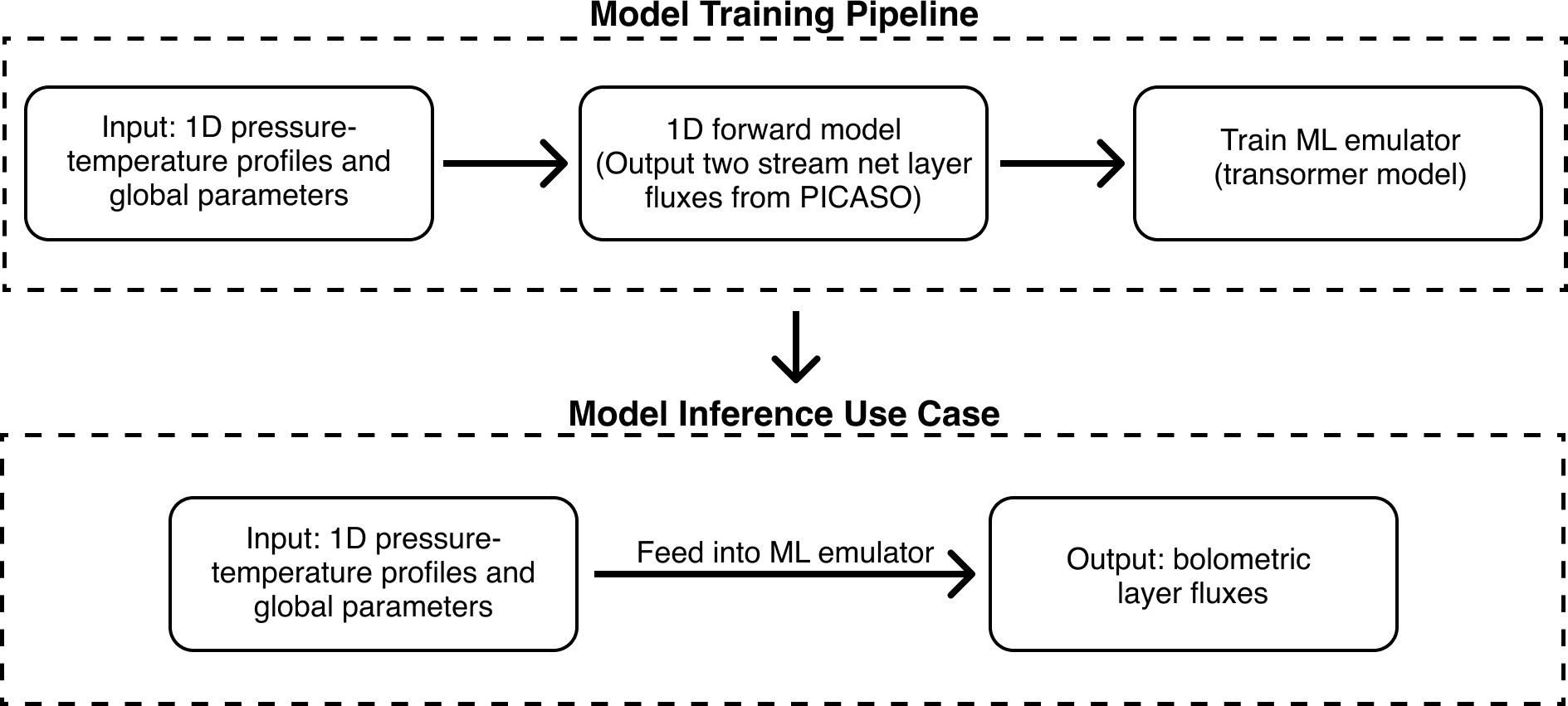}
\caption{The overall structure of the machine learning emulator pipeline presented in this work. First, 1D profiles are created based on randomly sampled parameterizations. Next, these profiles are processed with PICASO in order to solve the radiative transfer equations and determine the net thermal and starlight layer fluxes. Finally, these data are used to train an encoder-only transformer model that can predict radiative transfer fluxes based on an input pressure-temperature profile.}
\label{fig:flow}
\end{center}
\end{figure*}

The overall pipeline of this project is shown in Figure \ref{fig:flow}. To create a radiative transfer emulator, we first constructed a training set of randomly sampled 1D pressure-temperature profiles that cover a parameter space representative of hot-Jupiter atmospheres. These profiles were then fed through a traditional 1D forward model (PICASO; \citealt{Batalha2019, Mukherjee2023}), which calculated net layer fluxes using routines too computationally demanding for direct implementation in GCMs. Next, we trained the emulator on this corpus of data (after normalization). Finally, we characterized the model performance on a hold-out test set. %We compared several different emulator architectures on these data, such as varying transformer and recurrent neural network models.

\subsection{1D Pressure-Temperature Profiles}
We constructed a training dataset consisting of 1D atmospheric profiles. A broader underlying parameter space (e.g., the inclusion of clouds) would necessitate a larger training set. The pressure-temperature relation for each profile was set using a modified version of the \cite{line2013} equations for profiles in radiative equilibrium,

\[
T^4(\tau) = \frac{3T_{int}^4}{4} \left(\frac{2}{3} + \tau \right) + \frac{3T_{irr}^4}{4} (1 - \alpha) \xi_{\gamma_1} (\tau) + \frac{3T_{irr}^4}{4} \alpha \xi_{\gamma_2} (\tau)
\]

\noindent where $\xi$ is defined as 

\[
\xi_{\gamma_i} = \frac{2}{3} + \frac{2}{3\gamma_i} \left[1 + \left(\frac{\gamma_i \tau}{2} - 1\right)e^{-\gamma_i \tau} \right] + \frac{2\gamma_i}{3} \left(1 - \frac{\tau^2}{2} \right)E_2(\gamma_i \tau)
\]

\noindent and $\alpha$ is the partition between the two visible channels, $E_2$ is the second order exponential integral function, T$_{int}$ represents the bottom boundary heat flux, T$_{irr}$ is the irradiation temperature, $\tau$ is the pressure-dependent optical depth, and $\gamma_1$ and $\gamma_2$ are the ratios of the visible channel Planck mean opacities to the thermal opacity ($\kappa_{IR}$). Several additional parameters were introduced to (1) shift the entire pressure-temperature profile uniformly in temperature, and (2) adjust the thermal opacity following a prescribed pressure power law dependence following \cite{Robinson2012}, and (3) add a convective adjustment to a randomly selected fraction of the profiles. These adjustments were added in order to expand the parameter space from which we generated our 1D samples, and perturb the profiles from analytic radiative equilibrium. We adopted this parameterization with the goal of generating pressure-temperature profiles that form a super set of those seen in hot Jupiter atmospheres. In this work, we aim to model a canonical hot Jupiter. Extending the model to out of distribution atmospheres (e.g., sub-Neptunes) would require expanding the training data. However, this would not require fundamental changes to the underlying model.

\begin{figure*}
\begin{center}
\includegraphics[width=0.98\textwidth]{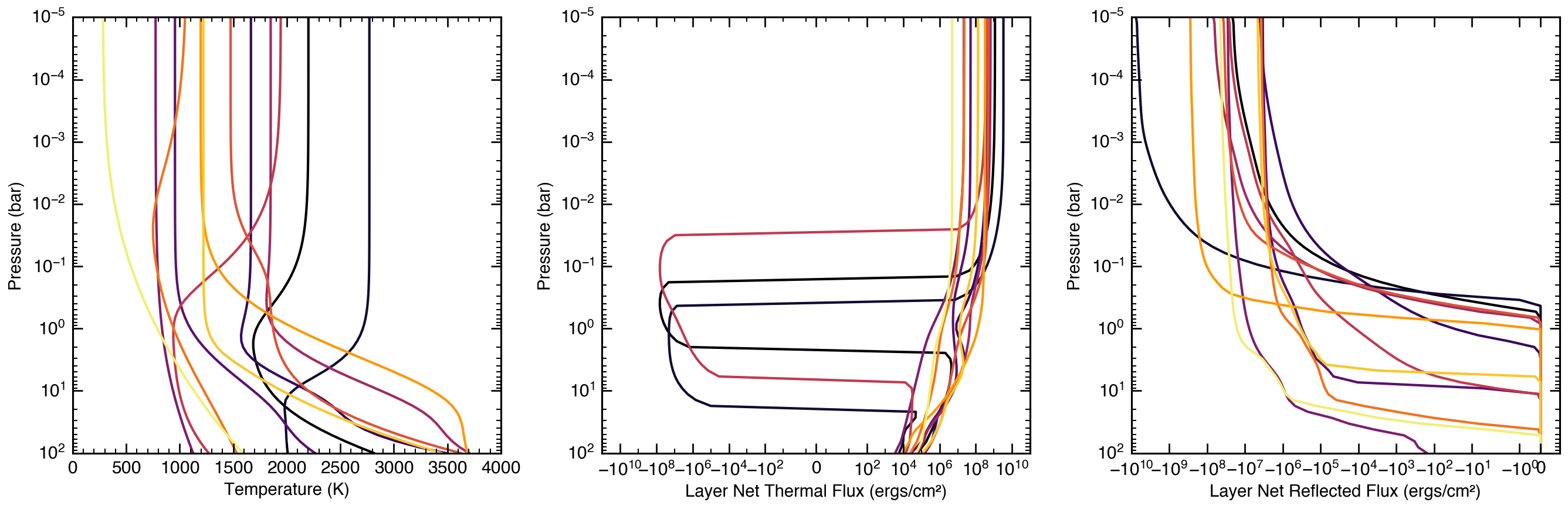}
\caption{Example 1D pressure-temperature profiles created using the parameterized relations from \cite{line2013}. These profiles were generated to be representative of a hot Jupiter atmosphere. We prioritize wide coverage of the parameter space, at the cost of simulating some unphysical atmospheres. The entire training set is composed of 2,000,000 such 1D profiles.}
\label{fig:pt}
\end{center}
\end{figure*}

The distribution parameters are listed in Table \ref{tab:config-params}. A sample of the pressure-temperature profiles and the associated thermal and scattered starlight layer fluxes is shown in Figure \ref{fig:pt}. We imposed hard limits that no profile could have negative temperatures or temperatures exceeding 4000 K, as this was the upper bound of temperatures for our correlated-k tables in PICASO. In total, we used 2,000,000 1D profiles. This number was chosen because this is where our model achieves the desired $\sim$1\% test set error. A larger training set would further decrease error, but with increasingly diminishing returns.

\begin{table*}[ht]
\centering
\label{tab:config-params}
\begin{tabular}{l r r l}
\hline
\hline
\multicolumn{4}{l}{\textbf{Uniform Distribution}} \\
\hline
\textbf{Parameter} & \textbf{min} & \textbf{max} & \textbf{Unit} \\
\hline
$\kappa_{IR}$ pressure power-law exponent & -0.5 & 1 & -- \\
$\log_{10}(\kappa_{\mathrm{IR}})$ & -2.5 & 2.5 & $\mathrm{m^2\,kg^{-1}}$ \\
$\log_{10}(\gamma_1)$ & -2 & 2 & -- \\
$\log_{10}(\gamma_2)$ & -2 & 2 & -- \\
$\alpha$ & 0.0 & 1.0 & -- \\
Temperature shift & -600 & 600 & K \\
\hline
\multicolumn{4}{l}{\textbf{Normal Distribution}} \\
\hline
\textbf{Parameter} & $\boldsymbol{\mu}$ & $\boldsymbol{\sigma}$ & \textbf{Unit} \\
\hline
$T_{\mathrm{int}}$ & 300 & 700 & K \\
$T_{\mathrm{irr}}$ & 1800 & 500 & K \\
Orbital separation modifier & 1.0 & 0.5 & -- \\
\hline
\multicolumn{4}{l}{\textbf{Other / fixed}} \\
\hline
Pressure range (bounds) & $10^{-5}$ & $10^{2}$ & bar \\
Convective adjustment & On: $1/3$ & Off: $2/3$ & -- \\
\hline
\hline
\end{tabular}
\caption{Distributions used to determine parameterized pressure-temperature profiles.}
\end{table*}

\subsection{PICASO Radiative Transfer}
After creating the 1D pressure-temperature profiles, we post-processed them with PICASO to determine the thermal and scattered starlight fluxes at each layer. These post-processed profiles make up all of our training, validation, and test set data. Although the incident stellar flux and thermal profile are physically coupled, this parameterization provides a realistic connection between pressure-temperature profiles and stellar heating while allowing for flexible sampling of parameter space. This flexibility is crucial for generating our training set, as atmospheric dynamics often drive vertical profiles in hot Jupiter atmospheres away from strict radiative convective equilibrium. Any radiative transfer emulator, in order to guarantee accuracy for real uses within a GCM, must be trained on equilibrium and non-equilibrium profiles.

For each 1D atmospheric profile, we calculated two separate sources of flux at each layer: thermal radiation and scattered starlight. To calculate the fluxes, PICASO also requires several other parameters, such as orbital separation, associated with each profile. All parameters are listed in Table \ref{tab:config-params}. To determine the incident downward stellar flux on the planetary atmosphere, we first calculated the orbital separation corresponding to a given irradiation temperature \citep{line2013}. This equilibrium flux was then multiplied by a factor randomly drawn from a normal distribution centered at unity with a standard deviation of 0.5, enabling us to train on a wider distribution of scattered starlight net layer fluxes.

Across all models, we assumed solar stellar metallicity, a stellar radius of 1 R${\odot}$, a planet radius of 1 R${J}$, a planet mass of 1 M$_{J}$, a solar C/O ratio (=0.458), and a PHOENIX model representing a 6000 K star \citep{Lodders2010, Husser2013}. For simplicity, we used a single fixed $\mu = 1$ (the cosine of the stellar zenith angle), i.e., normal incidence at the top of the atmosphere. As a result, the shortwave (scattered starlight) heating profiles in the training data implicitly assume purely vertical attenuation. Applying this emulator to a GCM with multidimensional geometry will therefore require extending the training set to include a range of $\mu$ values so that the model learns the $\tau/\mu$ slant-path dependence. However, the model was designed such that zenith angle, and other global variables like stellar radius or atmospheric metallicity, can be easily incorporated (see \ref{subsec:ml}). Gas opacities were computed within PICASO assuming chemical equilibrium with premixed 196-bin correlated-$k$ tables \citep{marley2021}. All models included the effects of Rayleigh scattering and collisionally induced absorption (H$_2$-H$_2$, H$_2$-He, H$_2$-H, H$_2$-CH$_4$, H$_2$-N$_2$) and other continuum sources such as H$_2^-$ and H$^-$.

\subsection{Data Normalization}
Before training, all data must be normalized to ensure that each feature (orbital separation, pressure, layer fluxes, etc.) is on a comparable numerical scale \citep{2021textbookzhang}. We found normalization method choice to be as crucial as model architecture in improving accuracy \citep{Ukkonen2021}. Because thermal emission and scattered starlight fluxes span many orders of magnitude, using an inappropriate normalization method can squash smaller values or obscure the underlying structure of the data. For each feature we tested a number of normalization methods and compared the resulting model test set error. Each normalization method was also based on the feature's physical meaning (e.g., keeping temperature positive) and its numerical range to preserve information crucial for accurate modeling. Table \ref{tab:norms} describes all methods used for this work. Many other methods were tested, including inter-quartile range, arctan compression, and sigmoid transformations. Each method maps values into bounded intervals, such as [-1, 1] or [0, 1], promoting numerical stability during training.

\begin{table*}[ht]
\centering
\begingroup
\setlength{\tabcolsep}{12pt}       % wider columns
\renewcommand{\arraystretch}{1.55} % taller rows
\small                              % smaller font

\begin{tabular}{lll}
\toprule
\textbf{Method} & \textbf{Equation} & \textbf{Notes} \\
\midrule

Log--min--max &
$\displaystyle \frac{\log_{10}(\max(x,\varepsilon)) - m}{M - m}$ &
\shortstack[l]{Maps positive data to $[0,1]$ in $\log_{10}$-space.\\
$m=\min(\log_{10}x)$, $M=\max(\log_{10}x)$.} \\
\addlinespace[8pt]

Z-score &
$\displaystyle \frac{x-\mu}{\sigma}$ &
\shortstack[l]{Standardization in linear space. Subtract the mean\\
and divide by the standard deviation.} \\
\addlinespace[8pt]

Log--standard &
$\displaystyle \frac{\log_{10}(\max(x,\varepsilon))-\mu_{\mathrm{log}}}{\sigma_{\mathrm{log}}}$ &
\shortstack[l]{Standardization in $\log_{10}$-space; $\mu_{\mathrm{log}}=\mathrm{mean}(\log_{10}x)$,\\
$\sigma_{\mathrm{log}}=\mathrm{std}(\log_{10}x)$. We use an $\varepsilon$ to avoid $\log 0$.} \\
\addlinespace[8pt]

Symmetric log &
\multicolumn{1}{c}{$\vcenter{\hbox{$
\begin{array}{@{}l@{\quad}l@{}}
\displaystyle \frac{x}{s\,\tau}, & |x|\le \tau\\[6pt]
\displaystyle \frac{\operatorname{sign}(x)\,\bigl[1+\log_{10}(|x|/\tau)\bigr]}{s}, & |x|>\tau
\end{array}
$}}$} &
\shortstack[l]{Logarithmic with a linear range near $0$ (set by $\tau$).\\
We choose $\tau$ as the first percentile of the data, and \\ set $s$ so normalized data fall in $[-1,1]$.} \\
\bottomrule
\end{tabular}

\endgroup
\caption{Normalization methods used for different features. All logarithms are base-10. The variable $x$ denotes the underlying data being normalized (e.g., layer-dependent pressure values). All normalization parameters (e.g., scaling values) were calculated on the training set exclusively to prevent data leakage. We used log--min--max for pressure, z-score for temperature, log-standard for orbital separation, and symmetric-log for the thermal and scattered-starlight net fluxes.}
\label{tab:norms}
\end{table*}

\subsection{Transformer Model and Machine Learning Background}\label{subsec:ml}
A transformer model is a type of deep learning architecture that uses a mechanism called attention to capture interdependencies between elements (tokens) in a sequence. In our case, each token corresponds to one atmospheric layer (pressure and temperature). The per-layer features are linearly projected to the d$_{\mathrm{model}}$-dimensional embedding. This architecture was first introduced by \cite{Vaswani2017}. Attention layers are coupled with feed-forward networks (FFNs, detailed below) that introduce non-linearity and expand model complexity.

Our model is bidirectional, as any element (layer) in the sequence (1D atmospheric profile) can influence any other element. This is distinct from temporal sequences, where later elements cannot influence earlier elements, and a technique referred to as causal masking is required. Finally, our model is encoder-only. Predicting the layer net fluxes from the input 1D profiles is a sequence regression task, and best suited to this class of model. In juxtaposition, auto-regressive tasks, such as language generation, are best implemented with encoder-decoder or decoder-only transformers.

\begin{figure*}
\begin{center}
\includegraphics[width=0.99\textwidth]{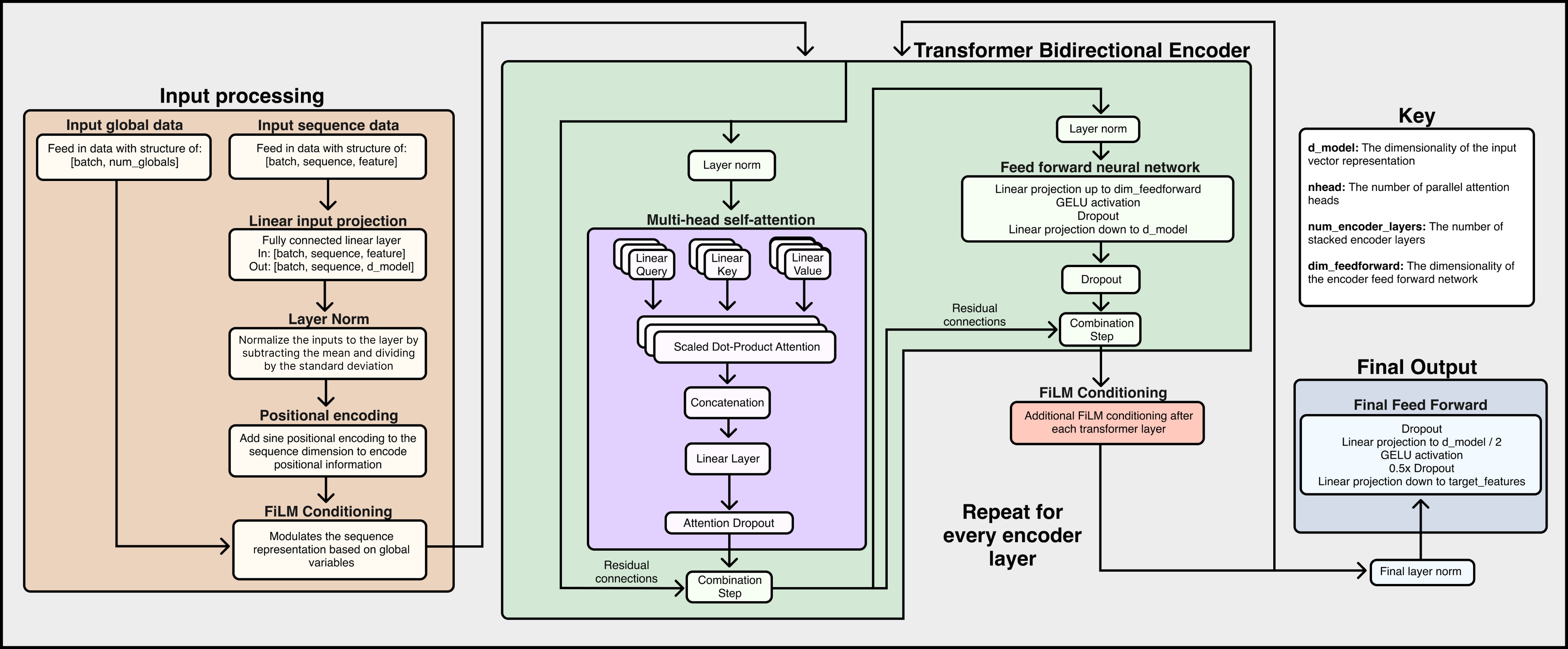}
\caption{A diagram showing how data flows through the transformer architecture implemented here. Data padding and masking are left out for simplicity. The model does implement this functionality, but it is not necessary for our constant-length sequences.}
\label{fig:fig}
\end{center}
\end{figure*}

The overall structure of our model---how data flows through our model and is modulated to predict the desired targets---is shown in Figure \ref{fig:fig}. This figure shows the various sub-processes required for the model as well. For example, dropout (randomly zeroing neuron outputs during training, for regularization) and residual connections (adding neuron input to neuron output to mitigate vanishing gradients). More background on the mechanisms within the model can be found in textbooks such as \cite{2025Ting} or \cite{2021textbookzhang}. Here, we only briefly describe the general structure of the model, and novel implementation. A list of all model and training parameters is shown in Table \ref{tab:params}. To implement this architecture, we used the PyTorch library \citep{Paszke2019}.

\begin{table*}[ht]
\centering
\begin{tabular}{@{}ll@{}}
\toprule
\multicolumn{2}{c}{\textbf{Model Parameters}} \\ \midrule
$d_{\mathrm{model}}$          & 256 \\ 
$n_{\text{head}}$             & 4 \\
Encoder Layers                & 4 \\
Feedforward Dimension         & 1024 \\
Dropout                       & 0.0 \\
Positional Encoding           & Sine \\
Activation                    & GELU \citep{Hendrycks2016}\\ \midrule
\multicolumn{2}{c}{\textbf{Training Configuration}} \\ \midrule
Input Variables               & pressure, temperature, orbital separation \\
Target Variables              & thermal net flux, scattered starlight net flux \\
Epochs                        & 300 \\
Loss Function                 & Mean Squared Error \\
Gradient Clip Value           & 2.0 \\
Optimizer                     & AdamW \citep{Loshchilov2017} \\
Initial Learning Rate                 & $1\times10^{-4}$ \\
Minimum Learning Rate          & $1\times10^{-8}$ \\
Weight Decay                   & $1\times10^{-5}$ \\ \bottomrule
\end{tabular}
\caption{Model parameters and training parameters.}\label{tab:params}
\end{table*}

The core transformer architecture is structured as a series of transformer encoder layers, each comprising multi-head self-attention (a method of determining how each element in the input sequences affects other elements, used to capture relationships within the input token sequences). The input tokens are linearly projected to $X \in \mathbb{R}^{d_{\mathrm{model}}}$ before adding positional information. Therefore each profile can be represented as $X\in\mathbb{R}^{L\times d_{\mathrm{model}}}$, where L is the number of layers per profile.

In order to incorporate positional information for each 1D input sequence, input features are linearly projected onto a vector of dimension $d_{\mathrm{model}}$ and standard sinusoidal positional encoding is applied,

\[
PE_{pos, 2i} = \sin\left(\frac{pos}{10000^{\frac{2i}{d_{\mathrm{model}}}}}\right)
\]
\[
PE_{pos, 2i+1} = \cos\left(\frac{pos}{10000^{\frac{2i}{d_{\mathrm{model}}}}}\right)
\]

\noindent where $pos$ is the index of each element in the input sequence, and i is the index within the resulting embedding vector and is an integer between 0 and $\frac{d_{\mathrm{model}}}{2} - 1$. Global parameters  (e.g., orbital separation) are incorporated  with sequential features using feature-wise linear modulation (FiLM; \citealt{Perez2017}) as

\[
[\gamma, \beta] = Wg + b,\quad \text{and}\quad \text{FiLM}(x,g) = (1 +\gamma )\odot x + \beta,
\]

\noindent where x are the sequential features, g are the global features, W and b are parameters learned during training and are used to determine the parameters $\gamma$ and $\beta$ from the global features g. During the forward FiLM pass, $\gamma$ and $\beta$ determine the affine transformation of x. This allows the model to incorporate the global features through a scaling and shifting of the sequential features. We apply an initial FiLM after the positional encoding, and after every encoder block, as shown by Figure \ref{fig:fig}.

We use standard scaled dot-product attention to calculate how each atmospheric level attends to every other atmospheric level. For each 1D profile, represented as $X\in\mathbb{R}^{L \times d_{\mathrm{model}}}$, we calculate query (Q$_i \in\mathbb{R}^{L \times d_{k}}$ ), key (K$_i\in\mathbb{R}^{L \times d_{k}}$), and value (V$_i\in\mathbb{R}^{L \times d_v}$) matrices as $Q_i=XW_{Q_i},\;K_i=XW_{K_i},\;V_i=XW_{V_i}$, using the learned linear projections W$_{Q_i}$, W$_{K_i}$ $\in\mathbb{R}^{d_{\mathrm{model}} \times d_{k}}$ and W$_{V_i}$ $\in\mathbb{R}^{d_{\mathrm{model}} \times d_{v}}$. We set d$_k$ = d$_v$ = d$_{\mathrm{model}}$/($n_{\text{head}}$). Because we use multi-head attention, the subscript i refers to the i-th attention head. Next, the attention weights (A$_i$) are defined as
\[
A_i = \operatorname{softmax}_{j}\!\left(\frac{Q_i K_i^\top}{\sqrt{d_k}}\right)
\]
with softmax defined as,
\[
S_i = \frac{Q_i K_i^\top}{\sqrt{d_k}},\qquad \mathrm{softmax((S_i)_{t,:})_j}= \frac{e^{(S_i)_{tj}}}{\sum_{m=1}^{L}e^{(S_i)_{tm}}}
\]

\noindent where t is the query index of S$_i$, j is the key index of S$_i$, and m is the summation index over keys for each query. Softmax ensures that the attention weights are between zero and one, and is applied row-wise, so that each row (query) sums to one across keys. Then, the weighted sum from each attention head is calculated as 
\[
H_i = A_iV_i
\]
\noindent Next, each output head ($H_i\in\mathbb{R}^{L \times d_{\mathrm{v}}}$) is concatenated and one final affine transformation is applied to produce the resulting output matrix $Y\in\mathbb{R}^{L \times d_{\mathrm{model}}}$. For simplicity, the batch dimension across all of these operations has been omitted, as well as any masking. Additionally, our final model did not include attention dropout.

Layer normalization is applied before each feed forward sub-layer and each multi-head attention sub-layer in our model, following \cite{Ba2016}. For every token fed into the model, we take the norm as
\[
y = \frac{x - \mu}{(\sigma^2 + \epsilon)^\frac{1}{2}} \odot  \gamma+\beta
\]
\noindent where $\gamma$ and $\beta$ are learned parameters. In this mechanism, the input token has the mean subtracted off, then is divided by the standard deviation (plus some small epsilon), and the result modified by the learned $\gamma$ and $\beta$. This standard approach stabilizes the training.

After multi-head self attention, the tokens are fed through feed forward neural networks comprised of an affine transformation (projecting up to $X \in \mathbb{R}^{L\times \mathrm{d_{FFN}}}$), a non-linear GELU activation function, dropout, and another affine transformation (down to  $X \in \mathbb{R}^{L\times \mathrm{d_{model}}}$). Although the attention mechanism is the 'heart' of transformer models, the majority of the parameters of the model come from these high-dimensionality feed-forward neural networks. For our activation function, we use GELU \citep{Hendrycks2016},
\[
\mathrm{GELU}(x)=x \times \Phi(x)
\]
\noindent the standard method within transformers. Tests with ReLU activations showed similar---though slightly worse---losses.

The final step in the model is an output head that takes the resulting vector of dimension d$_{\mathrm{model}}$, and applies a feed forward network, as shown in Figure \ref{fig:fig}. The output head projects from $X \in \mathbb{R}^{L\times \mathrm{d_{model}}}$ to $X \in \mathbb{R}^{L\times \mathrm{d_{targets}}}$). This is the final step necessary to return the predicted thermal and scattered starlight fluxes from the input sequence (and global data). We do not use any final output activation function.

\subsection{Training}\label{sec:training}
\begin{figure*}
\begin{center}
\includegraphics[width=0.99\textwidth]{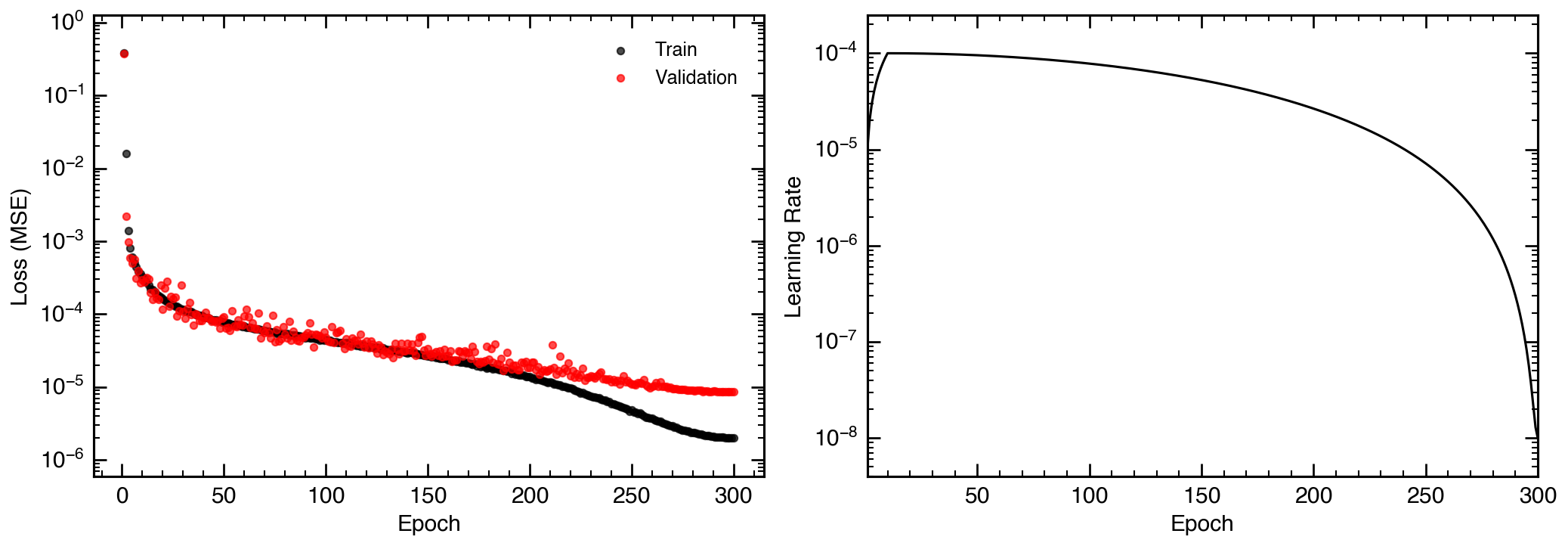}
\caption{Left: training loss vs. validation loss during model runtime. Right: the learning rate of the model. Near epoch 200, the model began overfitting, and the validation loss plateaued while the training loss continued to decrease, indicating moderate overfitting.}
\label{fig:training}
\end{center}
\end{figure*}

Figure \ref{fig:training} illustrates the progression of the model training process. The emulator was trained for 300 epochs. Hyperparameter tuning was conducted using the package Optuna \citep{Akiba2019}. We tested parameters including model dimensionality ($d_{\mathrm{model}}$, from 64 to 1024), feed-forward network dimension (from 128 to 4096), number of encoder layers (from 2 to 8), number of attention heads ($n_{\text{head}}$, from 2 to 8), activation functions (ReLU, GELU), as well as weight decay and dropout rates.

The data were divided into 70\% for training, 15\% for validation, and 15\% for testing. We used the Cosine scheduler \citep{Loshchilov2016} to automatically decrease the learning rate throughout training. Additionally, we used 10 warm-up epochs for initial training stability. The learning rate for our training is shown in Figure \ref{fig:training}. Our loss function (mean squared error) calculated the differences between the predicted and true layer fluxes at every sequence point and averaged the loss over the entire batch. Additionally, we used L2 weight decay \citep{2017arXiv171105101L}, and gradient clipping to ensure smooth training. With PyTorch compilation enabled, and automatic mixed precision, each epoch required $\sim$100 seconds on an NVIDIA A100 GPU for a total time of $\sim$8 hours.

\begin{figure*}
\begin{center}
\includegraphics[width=0.9\textwidth]{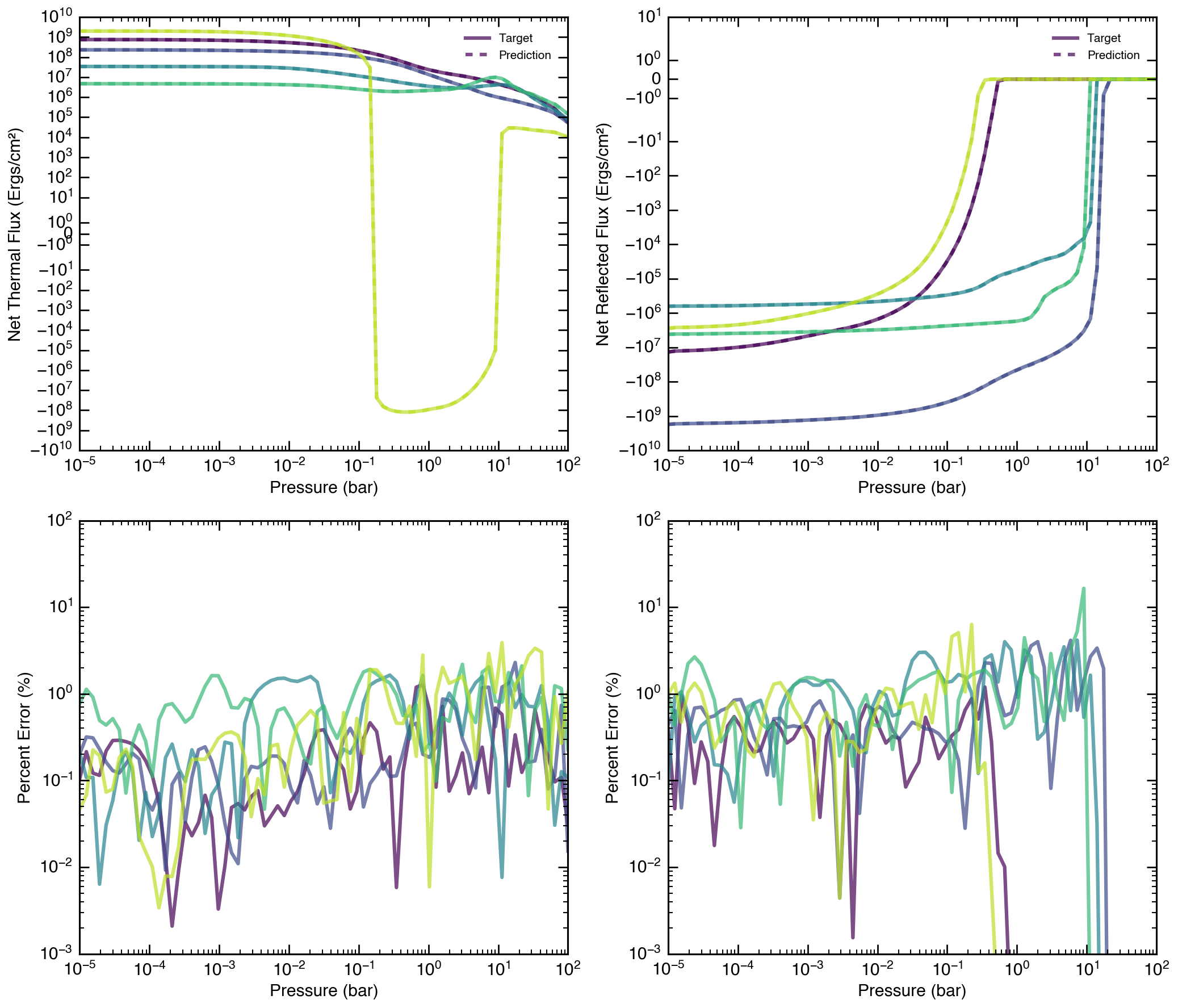}
\caption{The layer net fluxes (black) output from PICASO, compared to the values predicted by the transformer emulator (dashed red). Thermal fluxes and associated errors are shown in the top row, and scattered starlight fluxes and associated errors are shown in the bottom row. All profiles shown are drawn from the 15\% test set, that the model did not train on. Overall, we found that the emulator is within approximately 1\% of the true value.}
\label{fig:model}
\end{center}
\end{figure*}

\section{Results}\label{sec:results}
Figure \ref{fig:model} demonstrates the emulator's predictive performance on example profiles from the unseen 15\% test set. Overall, our emulator achieved errors for both the thermal and scattered starlight layer fluxes of $\sim$1\% relative to the PICASO outputs. The emulator was able to correctly model fluxes across all pressure ranges, even at transition points (such as thermal inversions), where the net layer fluxes change rapidly. To assess emulator accuracy across the entire test set (consisting of 300,000 profiles), we computed absolute and percent errors for each flux channel. Errors were defined with a small epsilon term to prevent large percent errors in deep layers where scattered starlight fluxes drop to zero,

\begin{equation}
\text{Signed Percent Error} = 100 \times \frac{\text{predicted} - \text{true}}{\max(\text{true}, \epsilon)}
\end{equation}

\begin{figure*}
\begin{center}
\includegraphics[width=0.8\textwidth]{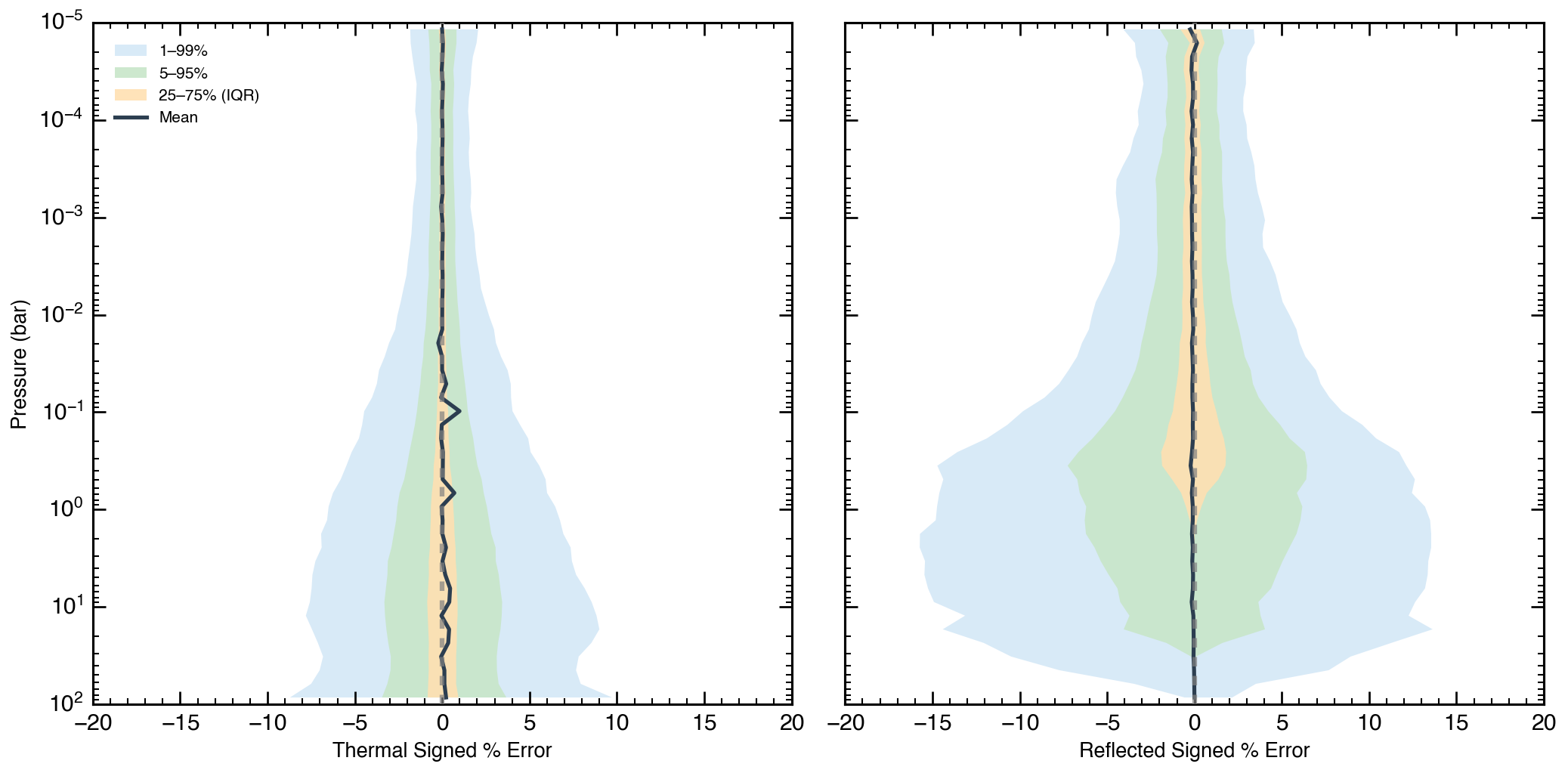}
\caption{The mean percent error for the thermal and scattered starlight layer net fluxes throughout the atmosphere, evaluated on the holdout test set. The solid black lines correspond to the error averaged across all layers. The colored bands show varying percentiles for the error distribution at each pressure level.}
\label{fig:error}
\end{center}
\end{figure*}

\noindent where $\epsilon$ was set to 1 erg cm$^{-2}$ s$^{-1}$. This small epsilon was used in order to avoid errors in the deeper atmosphere (where scattered starlight fluxes drop to 0) from dominating the analysis. Figure \ref{fig:error} shows the error of our surrogate model, compared to the PICASO data, for the entire test set. For the thermal channel, we found a mean (median) absolute percent error of 1.14\% (0.30\%). For scattered starlight fluxes, mean (median) absolute percent error was 1.26\% (0.45\%). For the thermal channel, we found a mean (median) signed percent error of 0.06\% (0.01\%). For scattered starlight fluxes, mean (median) signed percent error was -0.11\% (0.00\%). The signed errors for our model did not show a positive/negative bias. This is important for quantifying whether repeated radiative transfer calls, as required by a GCM, will inject/remove energy into the atmosphere.

We achieved the highest accuracy using a dropout rate of 0. The chosen model incorporates weight regularization and has a depth well-suited to the data, mitigating the need for dropout. While a more complex architecture with dropout could potentially improve accuracy further, it would also increase inference times. Therefore, we selected the parameters in Table \ref{tab:params} as the best balance between accuracy and speed.

\subsection{Inference Time}\label{sec:inference}
\begin{figure*}
\begin{center}
\includegraphics[width=0.5\textwidth]{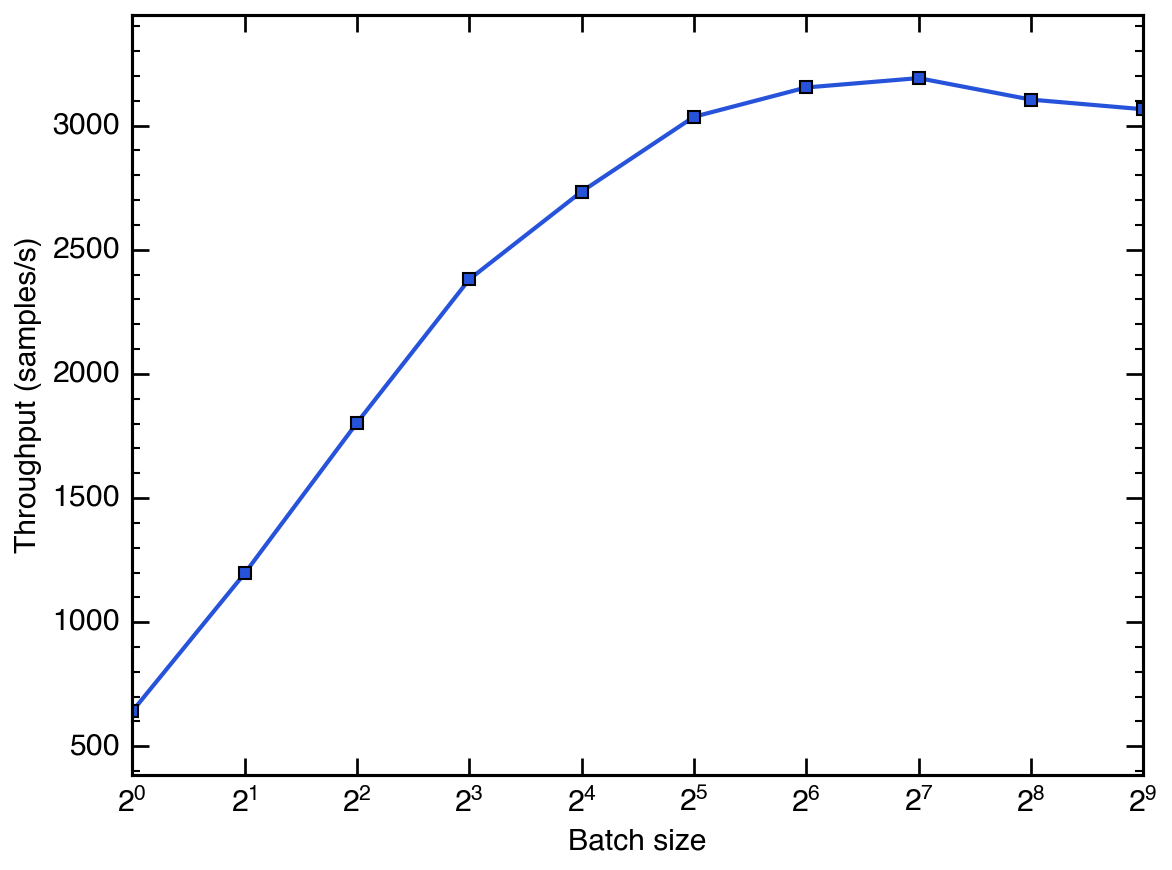}
\caption{Inference time for a full 75-layer sequence (thermal and scattered starlight channels) on an Apple M3 integrated GPU as a function of batch size. Larger batches amortize the fixed runtime overhead, reducing the average inference time required per sequence.}
\label{fig:benchmark}
\end{center}
\end{figure*}

On an Apple M3 integrated GPU, one stand-alone inference (full 75 layer sequence, both thermal and scattered starlight fluxes at every layer) took approximately 1.5 milliseconds. However, batching profile inferences and performing many calculations in parallel reduced the inference time to under 0.3 milliseconds per inference, as shown in Figure \ref{fig:benchmark}. These computations are for bolometric net flux, meaning that no wavelength-specific calculations are required (as the model was trained on integrated wavelength results). In comparison, using the Toon89 framework in PICASO with 196 wavelength bins for a 75 layer simulation would require approximately 400 milliseconds to run\footnote{Each wavelength bin must be calculated independently for both thermal and scattered starlight channels, and the computational cost scales as O(N), where N is the number of simulation vertical levels.}\citep{Toon1989, Rooney2023, Rooney2024}. However, the bolometric calculations required by GCMs in order to determine layer heating rates require dozens or hundreds of wavelength bins, meaning that the transformer approach provides orders of magnitude speed up compared to traditional methods. DISORT solutions are approximately three orders of magnitude slower than the Toon89 framework \citep{Rooney2023}, making them infeasible for GCM calculations. However, future work could use this source of radiative transfer training data to increase the emulator accuracy.

The speedups enabled by emulation would allow all necessary radiative transfer calculations required by a GCM to be executed on a single GPU or CPU over a timescale of several days. The model's efficiency could be further enhanced by reducing its parameter size or decreasing the vertical resolution of the training profiles. The input training profiles consist of 75 layers each, which is approximately double what is standard in most GCMs. However, transformers are resolution invariant (but not discretization-invariant) models \citep{Kovachki2021}, allowing greater possible flexibility for similar trained emulators. The model trained in this work was only trained on 75 layer profiles, and not tested on profiles of different vertical resolutions.

\section{Discussions and Conclusion}\label{sec:discussion}
Every machine learning emulator is essentially a black box, which necessitates careful validation to ensure that its outputs are correct. However, a radiative transfer emulator offers the advantage that its performance can easily be benchmarked against traditional 1D radiative transfer methods for any new test profile. The model presented here was trained on data corresponding to a single host star temperature, a single planet radius, and a solar metallicity atmosphere. Consequently, using it outside this parameter space would lead to dramatically increased errors.  Expanding the training set to encompass a broader range of input parameters would enhance the model's robustness and extend its applicability (see below).

Implicit in the emulator are the assumptions used to generate the training set. In this case, those assumptions include the two-stream approximation and the assumption of chemical equilibrium for the correlated‑k gas opacities. The model should not be interpreted as having learned the fundamental principles of radiative transfer; rather, its weights have been tuned to approximate the relationships present in the training data. Consequently, improving the accuracy of the training data would directly enhance the output model without increasing inference time.

In addition to the model presented above, we also explored a long short term memory (LSTM) recurrent neural network\footnote{A LSTM is a specific variant of a recurrent neural network, used for sequence learning tasks. For more information, see \cite{2021textbookzhang}.} similar to that of \cite{Ukkonen2021}, but found the transformer's flexibility in handling longer sequences while preserving accuracy to be superior --- particularly for non-local dependencies. The transformer model was particularly robust at capturing the non-locality of atmospheric radiative transfer, wherein each atmospheric layer simultaneously affects and is affected by all other layers.

\subsection{Future Work}\label{subsec:future_work}
In future work, we plan to maintain the encoder-only architecture while expanding our training data to encompass both pressure-temperature profiles and emission spectra. A natural extension would involve generating the training set using DISORT, as well as incorporating clouds and varying parameters such as planet characteristics, stellar properties, and gas opacities. Expanding the model to include varying chemistry assumptions, stellar temperatures, and planet types is feasible within our current setup. Our entire training dataset was less than 5 GB, and the model trained in under 8 hours. The A100 GPU that we used for training has 80 GB of memory, and data can also be streamed, rather than pre-loaded onto the GPU, if memory becomes a bottleneck. Additionally, different models can be created for different atmospheric populations, if training becomes prohibitively expensive. Therefore, this work can be expanded to allow the model to calculate radiative transfer for a larger parameter space of atmospheres.

A natural expansion for future work is to expand the sampled parameter space for our training data. Future models could vary stellar temperature, planet radius and mass, include clouds and/or hazes, and relax our assumption of chemical equilibrium. Although these changes would increase the size of the training dataset needed and model training time, it would not dramatically increase computational cost at inference time (the only increase in time coming from a larger model size). However, these expansions are outside the scope of this work.

In future work, we plan to incorporate this radiative transfer emulator within a 1D climate calculation, such as in PICASO. In following work, we plan to implement this module within a larger General Circulation Model, in order to greatly reduce the computational cost of the radiative transfer routines. Doing so would allow for an accuracy approaching the sophisticated 1D forward model used to train the model, while maintaining millisecond inference times. In the era of JWST and other next-gen telescopes, these improvements will be critical in interpreting observations and characterizing exoplanet atmospheres.

\section*{Acknowledgments}
IM would like to thank MC and ET for providing editorial suggestions, as well as CL, XZ, MD and MZ for helpful discussions. This research was carried out at the Jet Propulsion Laboratory, California Institute of Technology, under a contract with the National Aeronautics and Space Administration (80NM0018D0004).

\clearpage

%\bibliographystyle{aasjournal}
%\bibliography{bib}

\end{document}